# Acceptance of cybernetic avatars for capability enhancement: a large-scale survey

Laura Aymerich-Franch
Tarek Taha
Hiroko Kamide
Takahiro Miyashita
Hiroshi Ishiguro
Paolo Dario

**Authors**
Laura Aymerich-Franch[1], Tarek Taha[1], Hiroko Kamide[2], Takahiro Miyashita[3], Hiroshi Ishiguro[4], and Paolo Dario[1,5]

**Affiliations**
[1] Dubai Future Labs, Dubai Future Foundation, Dubai, United Arab Emirates
[2] Graduate School of Law, Kyoto University, Kyoto, Japan
[3] ATR (Advanced Telecommunications Research Institute International), Interaction Technology Bank, Keihanna Science City, Kyoto, Japan
[4] Department of Systems Innovation, Osaka University, Osaka, Japan and ATR Hiroshi Ishiguro Laboratories
[5] The BioRobotics Institute, Scuola Superiore Sant'Anna, Pisa, Italy

**Corresponding author**
Laura Aymerich-Franch
Dubai Future Labs, Dubai Future Foundation, Emirates Towers - Office Towers (Ground Level), Dubai, UAE
Laura.aymerich@dubaifuture.gov.ae

**Declaration of Interest statement**
The authors declare no competing interests

**Funding**

Research supported by Dubai Future Foundation and JST Moonshot R&D Grant Number JP-MJMS2011

# Acceptance of cybernetic avatars for capability enhancement: a large-scale survey

**Abstract**

Avatar embodiment experiences have the potential to enhance human capabilities by extending human senses, body, and mind. This study investigates social acceptance of robotic and virtual avatars as enablers of capability enhancement in six domains: identity exploration, well-being and behavioral transformation, expanded travel capabilities, expanded bodily and sensory abilities, cognitive augmentation, and immortality. We conducted a large-scale survey (n = 1001) in Dubai to explore acceptance of sixteen capability enhancement scenarios within these domains. The highest levels of agreement were observed for multilingual communication (77.5%) and learning capabilities (68.7%), followed by assisting individuals with reduced mobility (64.5%) and behavioral transformation (59.5%). Scenarios involving immortality through consciousness transfer received the least support (34.9%). These findings contribute to a deeper understanding of public attitudes toward avatar-based human enhancement and offer practical guidance for the responsible design, development, and integration of cybernetic avatars in the society, ensuring their societal acceptance and fostering a harmonious human-avatar coexistence.

## 1. Introduction

Cybernetic avatars are hybrid interaction robots or digital representations that can perform tasks on their own, while also being controlled or supervised by a human operator (Horikawa et al., 2023; Ishiguro, 2021; Ishiguro et al., 2025). Operators can interact socially through cybernetic avatars, control the avatar body, and communicate verbally through it (Aymerich-Franch et al., 2020; Dafarra et al., 2024; Hatada et al., 2024; Kishore et al., 2016). These entities are envisioned as a technology that can free humans from limitations of body, brain, space, and time (Ishiguro, 2021; Ishiguro et al., 2025). Robot avatars have primarily being explored in real-world applications as surrogate bodies for individuals with reduced mobility (Hatada et al., 2024) and for remote travel and physical interaction, such as virtual museum visits (Roussou et al., 2001). Additionally, virtual avatars are extensively used for real-world applications in gaming and interaction in social virtual reality applications, as well as for rehabilitation (Perez-Marcos, 2018)and for improving mental health (Aymerich-Franch, 2020a).

Beyond current applications, avatar embodiment experiences have the potential to enhance human capabilities by extending the human senses, body, and mind (Aymerich-Franch, 2020b; Biocca, 1992; Biocca & Delaney, 1995; Horikawa et al., 2023; Ishiguro et al., 2025). Capability augmentation has been explored among others through multiple avatar control by a single operator (Glas et al., 2012; Hatano et al., 2023; Kishore et al., 2016), modified avatar body structures (Steptoe et al., 2013; Won et al., 2015) and modified avatar appearances (Aymerich-Franch et al., 2014; Rosenberg et al., 2013; Yee & Bailenson, 2007).

In Aymerich-Franch (2020b), four key directions through which avatar embodiment systems can enhance human capabilities were identified: embodiment of a new self, expanded traveling capabilities, expanded body capabilities, and immortality. Embodiment of a new self is associated with the liberating potential of becoming someone different through an avatar. Avatar embodiment experiences allow users to express their authentic selves without the constraints of physical identity or social norms, offering a safe space for self-exploration and identity transformation (Spears & Lea, 1994). For instance, users may experience reduced anxiety and inhibition by disassociating from their real-world identity, thereby improving social interaction (Suler, 2004). These transformations range from subtle changes such as eye or hair color to drastic alterations in gender, age, skin color, or even species (Ahn et al., 2016; Gonzalez-Liencres et al., 2020; Hershfield et al., 2011; Johnston et al., 2023; Peck et al., 2013). It is also known that embodiment in avatars with different physical traits can significantly influence user behavior, a phenomenon known as the Proteus Effect (Yee & Bailenson, 2007). This effect arises from the association of avatar traits with behavioral expectations, shaping how users act within virtual environments. Expanded body capabilities are related to optimized body designs for efficiency, extended bodies for people with reduced mobility, and new sensory and motor opportunities. A physical avatar can possess capabilities beyond those of a typical human body, such as increased strength, the ability to fly or remain underwater for extended periods, and resistance to extreme environmental conditions (Won et al., 2015). These capabilities have practical applications in hazardous contexts, such as disaster response, and could also support individuals with disabilities in everyday tasks (Aymerich-Franch, 2020b; Ishiguro, 2021), and make them able to work remotely, thus restoring their inclusion in the job market (Hatada et al., 2024). On the other hand, digital avatars, being non-corporeal, are free from physical constraints like gravity and perishability. It is known that users are capable of experiencing body ownership of an avatar body even when avatars do not resemble their real appearance(Ahn et al., 2016; Aymerich-Franch et al., 2017), and can learn to control novel structures such as tails or additional limbs (Steptoe et al., 2013; Won et al., 2015).

Additionally, in Aymerich-Franch (2020a), avatar embodiment is highlighted as a technology with strong potential for behavioral transformation and enhancing mental health. Avatar embodiment has demonstrated therapeutic benefits for reducing anxiety, social anxiety, public speaking fear, depression, and self-criticism (Aymerich-Franch et al., 2014; Aymerich-Franch & Bailenson, 2014; Falconer et al., 2016; Osimo et al., 2015; Slater et al., 2019). It also shows promise for reducing racial bias (Groom et al., 2009; Peck et al., 2013), promoting empathy and prosocial behavior (Rosenberg et al., 2013), and encouraging pro-environmental attitudes (Ahn et al., 2016; Bailey et al., 2015). Moreover, avatar embodiment technologies offer potential for neurorehabilitation (Perez-Marcos, 2018). Embodiment experiences have also been used to promote empathy and reduce aggression in contexts like domestic violence(Gonzalez-Liencres et al., 2020; Johnston et al., 2023), and have also been linked to health behavior change through the use of doppelgängers in exercise and eating habit interventions (Fox et al., 2009; Fox & Bailenson, 2009).

Lastly, avatar embodiment experiences have also shown that changes can also occur in cognitive processing and, specifically, executive functioning. In a virtual reality experiment, participants embodied in an Einstein-looking avatar, performed better on a cognitive task than the normal body, with the improvement greatest for those with low self-esteem, which shows that avatars could also be used to enhance executive functioning (Banakou et al., 2018).

To explore social acceptance of robotic and virtual avatars as enablers of capability enhancement, we conducted a large-scale survey with over 1,000 participants in Dubai. Dubai's demographically diverse environment provides a rare opportunity to examine our object of study across highly varied geographical backgrounds within a single setting. This study capitalizes on the Emirate's cosmopolitan character to present a uniquely global perspective on the social acceptance of this emerging technology.

## 2. Methodology

### 2.1. Survey content and distribution

We extracted the various activities associated with the four categories of enhanced capabilities highlighted by (Aymerich-Franch, 2020b), namely, embodiment of a new self, expanded traveling capabilities, expanded body capabilities, and immortality. We additionally added a series of new tasks related to expanded cognitive abilities (Osimo et al., 2015), and with well-being and behavioral transformation (Aymerich-Franch, 2020a), which resulted in the following six domains of avatar-based capability enhancement:

- Identity exploration and appearance customization (i.e., creating new identities and experiencing becoming someone different);

- Customizing one's appearance such as eye, hair, skin color, or body shape to adopt new human or non-human forms;
- Expanded travel capabilities, referring to remote and immersive access to locations beyond one's immediate environment (i.e., traveling to remote locations such as other planets or countries; exploring fantasy worlds; flying; or staying underwater for extended periods);
- Expanded body and sensory capabilities, which includes physical augmentation and assistive functions (i.e., gaining additional limbs or tails to improve physical performance; controlling multiple avatars to overcome the limitations of a single body; experiencing new senses; or using avatars as complementary bodies for individuals with reduced mobility);
- Expanded cognitive abilities, involving improved interaction and mental processing (i.e., interacting with people in other languages; adopting new ways of learning or reasoning);
- Well-being and behavioral transformation, focused on psychological, emotional, and social development (i.e., using avatars for behavioral transformation and self-improvement, such as reducing racial bias or promoting prosocial behavior; supporting mental health treatments for anxiety, depression, or eating disorders; or serving as a pain distraction method during medical procedures); and
- Immortality, which explores the speculative continuation of the self through digital means (i.e., transferring consciousness to an avatar to achieve immortality).

These six categories formed the structure of our survey, with each of the 16 enhancement scenarios aligned accordingly. A full list of items can be found in the Supplementary Material (SM – 1).

We asked the participants whether, in the ideal future society they envisioned for Dubai, they would like to have robot and digital avatars for enhancing the 16 different capabilities (listed in the Sup. Material). The participants rated each item on a scale of *disagree – neutral – agree*. We also included an open-ended question inviting participants to imagine an ideal robot or virtual avatar and describe which human capability they would most like it to enhance.

Regarding demographics, we collected information on nationality, gender, age, occupation, time living in the UAE, born in the UAE, level of studies, income level, religious beliefs, background in Computer Science, level of programming skills, experience using robots, experience interacting with social robots and avatars, and collectivism – individualism orientation (Triandis & Gelfand, 1998). We additionally collected information related to relationship with robots. In particular, we included general level of interest in scientific discoveries and technological developments, general view of robots and attitudes towards robots scale (items based on (Aymerich-Franch &

Gómez, 2024) and (European Commission, 2012), and fear of robots scale (items based on (Aymerich-Franch & Gómez, 2024; Liang & Lee, 2017). We additionally added two attention check questions among the previous questions (e.g., "I am paying attention to the survey, select "agree"). At the start of the survey, we provided a visual introduction to the concept of robot avatars, which were defined as "hybrid interaction robots that combine autonomous capabilities with teleoperated control. They can interact socially and perform tasks on their own, while also being controlled by a human operator for real-time interaction"[*].

The survey was outsourced to an external market research firm, which was responsible for recruiting participants and administering the survey through its platform. Participants were compensated with points, which could be accumulated and exchanged for rewards. Data collection took place between December 2024 and February 2025. Upon completion of the survey, the market research firm provided the raw data in Excel format, which we used for subsequent analysis.

## 2.2. Participants

The survey was administered in English and it was open to participants aged 18 or older, residents of Dubai, from any of the 6 community clusters listed in Table 1.

The survey was initiated by 4,000 participants. Of these, 1,542 were screened out due to ineligibility, 883 were excluded after quota targets were reached, 198 did not finalize the survey, and 376 were removed through quality control checks. These measures helped ensure a high-quality and demographically balanced dataset for analysis. The final sample consisted of 1,001 participants. The study received ethical approval from the [institution hidden for peer review] Ethics Committee.

### 2.3.1. Community Clusters Distribution

The population size of the Emirate of Dubai is estimated at 3.8 million, having the highest population among the emirates in the UAE. Of them, 31.4% are females and 68.6% are males. The majority of the residents (58.49%) are aged between 25 to 44 years.

Over 200 nationalities live and work in the UAE (UAE Ministry of Foreign Affairs, 2024). Given that the objective of the study was to capture perspectives from all major communities residing in Dubai, and act at the same time as a rather than to obtain a statistically representative sample of the general population, we employed a structured sampling strategy. Specifically, a stratified sampling method was used, based on gender and community affiliation, to ensure balanced representation.

---

[*] The participants completed additional blocks of questions related to acceptance of robot avatars for the service sector which are not reported here.

For that, we identified the expat communities with 10,000 residents or more in the UAE (Wikipedia, 2024). A total of 35 countries and areas were identified. We then classified the countries by geographical region following the classification provided by the Statistics Division of the United Nations for statistical use (UNSD, 2024) to get an understanding of the geographical distribution of the represented groups. The initial classification resulted in countries and areas in the following regions: Eastern, South-Eastern, Southern, and Western Asia; Northern and Sub-Saharan Africa; Eastern, Northern, Southern and Western Europe; Northern America; Australia and NZ. We then further organized the countries and regions into six main community clusters that best reflect Dubai's social composition: Emiratis (UAE citizens), a Middle East cluster (neighboring countries), a South Asian cluster (representing the largest expatriate group, comprising nearly 60% of the UAE population), and three additional clusters for 'Other Asian', 'Other African', and 'Western' (Table 1).

| Cluster | Countries / Areas | Participants (N) |
|---|---|---|
| Emirati | United Arab Emirates | 170 (83 males, 87 females) |
| Middle East | Egypt, Iraq, Jordan, Lebanon, Palestine, Syria, Saudi Arabia, Türkiye, Qatar, Kuwait, Bahrain, Oman | 166 (84 males, 82 females) |
| Southern Asia | India, Pakistan, Bangladesh, Nepal, Sri Lanka | 173 (89 males, 84 females) |
| Other Asia | Philippines, China, Indonesia, Japan, Taiwan, Malaysia, Myanmar, Singapore, Kazakhstan, Uzbekistan, Armenia, Azerbaijan | 174 (81 males, 93 females) |
| Western | European Union (Austria, Bulgaria, Croatia, Cyprus, Finland, France, Germany, Greece, Ireland, Italy, Lithuania, Poland, Portugal, Romania, Sweden), United Kingdom and overseas territories, United States and territories, Russia, Canada, Australia, Switzerland, Belarus, Bosnia Herz, Ukraine, Serbia, New Zealand, Albania, Peru, Cuba, Panama, Dominican Rep. | 161 (85 males, 76 females) |
| Other Africa | Ethiopia, Kenya, Benin, Cameroon, Ghana, Central Africa Rep, Congo, Cote Ivoire, Eritrea, Gambia, Ghana, Guinea, | 157 (76 males, 81 females) |

| | Nigeria, Rwanda, Sierra Leone, Tanzania, Togo, Uganda, Zimbabwe, Comoros | |

**Table 1.** *Clusters by country / areas, and sample representation in the large-scale survey.*

## 3. Results

The final sample consisted of 1,001 participants, 503 females and 498 males. Participants were distributed across six community clusters, each comprising between 157 and 174 individuals: Emirati (170), Middle East (166), Southern Asia (173), Other Asia (174), Western (161), and Other Africa (157). Of the participants, 4.1% were between 18–24 years old, 38.0% were 25–34, 38.4% were 35–44, and 19.6% were 45 or older. Most respondents (42%) had lived in Dubai between 4–10 years, with 24.1% born in the UAE. Over half (53.6%) indicated a high interest in scientific discoveries and technological developments, and the majority expressed a positive general view of robots, with 53.7% fairly positive and 31.3% very positive. Fear of robots was moderate overall: 30.9% reported no fear, while 53.2% reported slightly or moderate fear. Most participants had occasionally interacted with robots (58.2%), and 54.1% had occasionally interacted with avatars.

Overall, acceptance of cybernetic avatars for capability enhancement was found to be generally favorable, with most categories receiving approval rates exceeding 40%. This suggests a broad willingness among participants to integrate avatar embodiment systems into future society, particularly in Dubai. However, variations exist in the degree of acceptance depending on the specific application.

Among the various applications considered, three categories received the highest levels of acceptance: employing avatars for being able to speak with people in other languages, with 77.5% of participants expressing agreement, adopting new ways of learning or reasoning (68.7%), and using avatars as complementary bodies for individuals with reduced mobility (64.5%). Conversely, the concept of transferring consciousness to an avatar to achieve immortality received the lowest level of acceptance, with only 34.9% in favor, while 35.2% actively opposed it. This suggests that ethical and existential concerns may hinder the broad adoption of this specific application.

To further explore these findings, we categorized the results into six primary areas of capability enhancement: (1) identity exploration, (2) well-being and behavioral transformation, (3) expanded travel capabilities, (4) expanded body and sensory capabilities, (5) expanded cognitive abilities, and (6) immortality.

Participants demonstrated moderate to strong acceptance of avatars as a means of identity exploration and personal transformation. Creating new identities and experiencing becoming someone different was agreed upon by 40.8% of participants, while customizing one's appearance, including changes to race, gender, and non-human forms, was supported by 42.3%. These findings indicate a substantial recognition of avatars as tools for psychological and social adaptation.

Using avatars for behavioral transformation and self-improvement, such as reducing racial bias and promoting pro-environmental behaviors, received strong approval at 59.5%. Avatars were also seen as beneficial for mental health applications, such as treating phobias, anxiety-related disorders, and eating disorders, with 52.4% in agreement. Additionally, using an avatar as a pain distraction method during medical operations was widely accepted, with 50.2% in agreement. These findings suggest that avatars are perceived as valuable tools for mental health and well-being.

Acceptance of avatars as a means of overcoming physical travel limitations was generally high. Traveling remotely through a robot avatar, including visits to other planets or distant countries, was supported by 55.4% of participants. Similarly, exploring fantasy worlds through digital avatars received the same level of approval (55.4%). The ability to fly using a robot avatar was endorsed by 42.0% of participants, while staying underwater for extended periods was supported by 47.7%. These results suggest that avatars are seen as effective alternatives for both physical and virtual exploration.

Enhancing physical abilities through avatars also received notable levels of approval. The idea of adding extra limbs to enhance physical performance, such as completing tasks more efficiently, was approved by 52.7% of respondents. Controlling multiple avatars to overcome the limitations of a single physical body was agreed upon by 44.7%. The strongest support in this category was for using robot avatars as complementary bodies for individuals with reduced mobility, with 64.5% of participants favoring this application. Additionally, experiencing new senses was supported by 58.1%. This suggests that while extreme modifications to the human form may be met with some hesitation, practical applications for accessibility and efficiency are widely embraced.

The use of avatars to expand cognitive abilities received strong support. The ability to interact with people in other languages was widely accepted, with 77.5% approval, while having new ways of learning or reasoning was supported by 68.7%. These results highlight the role of avatars in enhancing cognitive and communication capabilities.

The notion of achieving immortality by transferring consciousness to an avatar was the least accepted enhancement, with only 34.9% approval. A significant portion of participants (35.2%) explicitly disagreed with this application, while 30.0% remained neutral. This finding underscores the ethical, philosophical, and possibly religious concerns associated with the idea of digital immortality, making it a more controversial and less widely accepted concept compared to other forms of avatar embodiment.

Figure 1 shows the percentage of acceptance for each of the items.

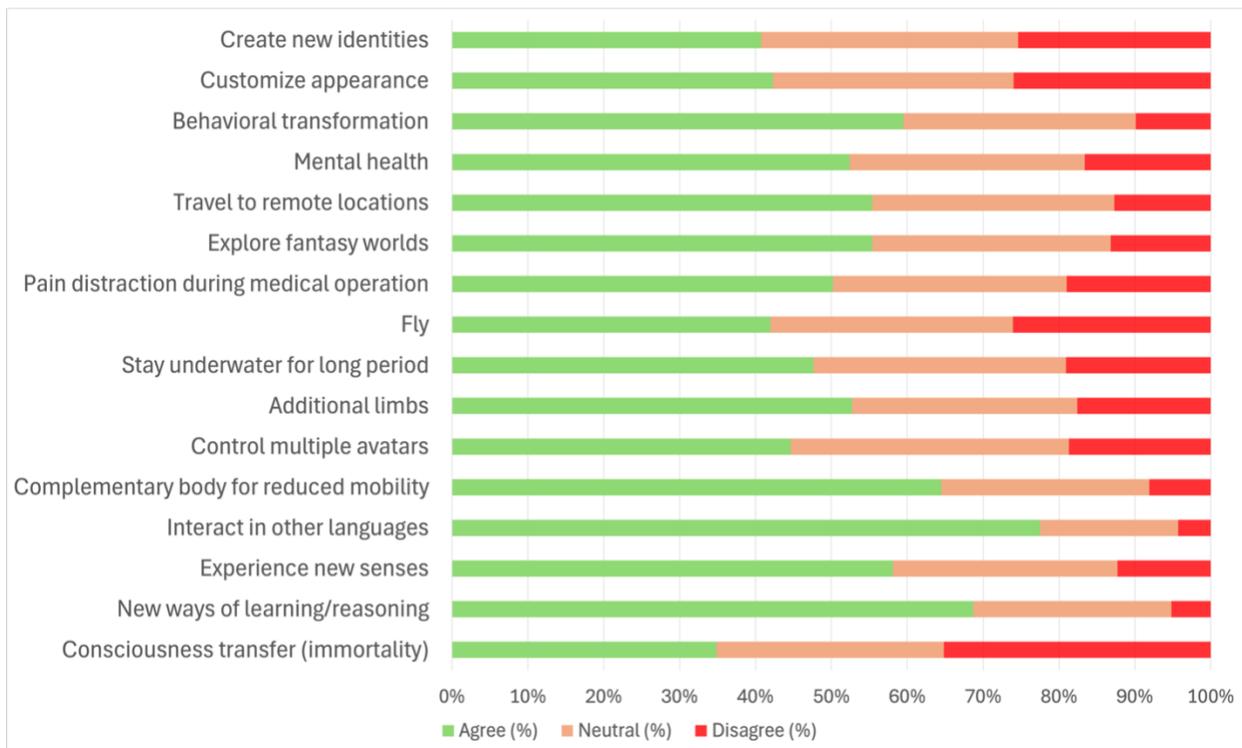

*Figure 1. Acceptance of cybernetic avatars for enhanced capabilities*

### 3.1. Open-ended answer analysis

A thematic analysis was manually conducted by one of the researchers on the open-ended responses to the question about which human capability participants would most like to enhance if they could create an ideal robot or virtual avatar. Responses were first screened for relevance and quality prior to analysis; 111 non-responses and 23 answers consisting only of "yes" were excluded. The remaining responses were then coded iteratively to identify key themes, with low-quality (e.g., single non-related word or symbol such a percentage) or off-topic answers skipped during coding. Coding continued until thematic saturation was reached (after response num. 360), meaning no new themes or insights emerged from additional data.

In line with the results of the previous questions, we identified communication abilities, augmented body capabilities, and cognitive abilities emerging as the most dominant themes. Other, less frequent but noteworthy themes included sensory abilities, mental health and emotional support, work-related assistance, superhero-inspired capabilities and altruistic intentions, and no wish for enhancement. A selection of representative quotes by category can be found in the Supplementary Material (SM – 2).

An additional independent analysis was performed using AI-assisted methods to identify recurring semantic patterns and identify key themes by grouping responses based on keyword similarity and latent linguistic features using lexical matching and semantic proximity. The AI analysis produced similar dominant themes to those identified by the human coder, which strengthened the reliability of our themes: Communication & Language; Cognitive Enhancement; Physical or Sensory Abilities; Creativity & Imagination; and Emotional Intelligence & Empathy.

The final themes are further developed in the following sections based on the human coder analysis.

*3.1.1. Communication Abilities*

The most prominent theme concerned enhancing communication and empathy. Participants envisioned robotic and digital avatars capable of transcending linguistic barriers, helping humans to connect more effectively across cultures, with multi-lingual capabilities, as well as capable of interpreting emotional states and foster understanding and inclusivity in diverse settings (e.g. "I would like the robot or virtual avatar to enhance communication skills, such as translating languages in real time, understanding emotions…").

*3.1.2. Cognitive Abilities*

The second theme focused on amplifying intellectual and creative capacities. Participants expressed a strong interest in robotic and digital avatars that could boost creativity, memory, decision-making, problem-solving, analyze data, time management skills, take better decisions, adapt quickly to new situations, and learning (e.g., "If I could create the ideal robot or virtual avatar, I would want it to enhance human creativity. By amplifying creative thinking, this avatar could help people explore new ideas, solve problems in innovative ways, and unlock their full imaginative potential…)".

*3.1.3. Body Capabilities*

A third main theme was related to physical enhancement. Participants imagined robots or avatars with superior strength, agility, resilience, and mobility, capable of performing physically demanding or dangerous tasks. The imagined capabilities ranged from lifting heavy objects and moving quickly, to enduring extreme conditions or accessing hazardous environments, such as rescuing people from fires (e.g., "Super strength that can easily carry heavy objects."). Several participants also emphasized inclusive physical augmentation, such as enhancing eyesight or mobility for people with disabilities. These responses indicate that physical enhancement was understood both in terms of power and protection, and in terms of accessibility and support.

*3.1.4. Other Capabilities*

A smaller subset of participants mentioned about providing avatars with the human senses of touch, smell, and sight. Other participants showed interest in having avatars capable of providing emotional well-being and psychological care. A few participants pointed at enhancement with work and daily productivity, envisioning robots that could complete household chores, or handle professional tasks. Some participants drew inspiration from superheroes or celebrities, imagining robots with superhuman traits or altruistic missions, referring among others to having avatars that can do what Spiderman or BB8 from Star Wars could do. Others mentioned robots that could undertake humanitarian causes.

*3.1.5 No Wish for Enhanced Capabilities through Avatars*

A final group of respondents expressed no wish to enhance human capabilities through robotic and digital avatars.

**4. Discussion**

This study presents the first large-scale investigation into public acceptance of avatar embodiment for different types of human capability enhancement. Participants in Dubai evaluated 16 enhancement scenarios mapped to six functional domains. Overall, we observed broad willingness to adopt cybernetic avatars, with the majority of scenarios receiving more than 40% acceptance. The highest levels of agreement were found for practical and socially meaningful applications, such as multilingual communication and support for people with reduced mobility. In contrast, more speculative or philosophically complex scenarios, such as transferring consciousness to achieve immortality, received the least support.

These findings point to a societal preference for avatar applications that offer direct and tangible benefits to daily life, accessibility, communication, and health. The relatively high acceptance of mental health and behavioral transformation scenarios underscores the emerging legitimacy of virtual environments as therapeutic and enabling tools for healthcare-related applications.

Cybernetic avatars are progressively being deployed in Japan and potentially will also expand in other societies. In Japan in particular, there is increasing interest in this technology which is receiving important R&D attention under the Moonshot program, which envisions the "Development of technologies that will allow anyone willing to augment their physical, cognitive, and perceptional capabilities to the top level, and spread of a new lifestyle that will be welcomed by society, by 2050." As part of this initiative, we are also actively deploying cybernetic avatars in Dubai to explore different scenarios and functions as part of the project. By gathering first-hand insights from Dubai citizens, this work provides early perspectives on acceptance and application potential within Dubai's multicultural society. These insights are not only vital for informing local avatar deployment strategies but also serve as foundational feedback to shape avatar systems under the Moonshot framework. By incorporating stakeholder input into the design and development process, the aim is to ensure that future avatars align with the cultural, social, and practical expectations of users in diverse environments like Dubai (Kamide et al., 2025) —ultimately enhancing acceptance locally and serving as a reference for global deployment.

The findings of this study offer concrete insights for guiding the responsible development and deployment of avatar technologies. By identifying which capability enhancement scenarios are more socially accepted, the results help prioritize applications with strong public support—such as multilingual communication, cognitive augmentation, and support for individuals with reduced mobility—for near-term implementation. The study also underscores the importance of designing avatar systems that reflect the cultural, social, and practical expectations of diverse user populations, particularly in multicultural contexts like Dubai. Furthermore, the nuanced responses across different domains highlight the need for early stakeholder engagement and public dialogue to address ethical concerns and foster widespread societal acceptance of cybernetic avatars.

Further work should explore longitudinal changes in public perception as avatar technologies become more integrated into everyday life. Future studies could also investigate acceptance in different cultural contexts, expand the set of capability enhancement categories, and assess actual user experiences beyond self-reported intention through experimental research and deployments in real scenarios. For instance, multilingual capabilities and cross-cultural awareness emerged as critical needs to be performed by avatars, particularly relevant in Dubai's highly multicultural environment. It would be interesting to examine which capabilities would be prioritized if the study were replicated across different geographical contexts. Additionally, integrating qualitative insights would enrich our understanding of the motivations, concerns, and ethical reflections behind acceptance judgments. As the field evolves, multidisciplinary collaboration will be essential for guiding the responsible and equitable integration of avatars into society.

UNSD. (2024). *Methodology. Countries or areas / geographical regions*. Https://Unstats.Un.Org/Unsd/Methodology/M49.

Wikipedia. (2024). *Demographics of the United Arab Emirates*. Https://En.Wikipedia.Org/Wiki/Demographics_of_the_United_Arab_Emirates.

Won, A. S., Bailenson, J., Lee, J., & Lanier, J. (2015). Homuncular Flexibility in Virtual Reality. *Journal of Computer-Mediated Communication*, *20*(3). https://doi.org/10.1111/jcc4.12107

Yee, N., & Bailenson, J. (2007). The proteus effect: The effect of transformed self-representation on behavior. *Human Communication Research*, *33*(3). https://doi.org/10.1111/j.1468-2958.2007.00299.x

**Supplementary Material for the article:**
**"Acceptance of cybernetic avatars for capability enhancement: a large-scale survey"**

**SM – 1**

Full list of items included in the survey:

*Imagine a future society in which robot and virtual avatars are introduced in Dubai as a way to enhance human capabilities. Everyone would be able to use these avatars for functions such as expanding body and travel capabilities, among others.*

*In the ideal future society you envision for Dubai, robot and virtual avatars should be used...*

*Identity exploration*
• *To create new identities and experience becoming someone different*
• *To customize oneself by changing appearance (eye, hair, skin color, or body shape) for creating new human and non-human appearances*

*Well-being and behavioral transformation*
• *For behavioral transformation and self-improvement (reducing racial bias, promoting pro-environmental and pro-social behavior)*
• *For mental health applications (treating phobias, anxiety-related disorders, substance-related disorders, or eating disorders*
• *During a medical operation, as a pain distraction method*

*Expanded travel capabilities*
• *To travel to remote locations (such as other planets or other countries)*
• *To explore fantasy worlds*
• *To fly*
• *To stay underwater for a long period*

*Expanded body and sensory capabilities*
• *To get additional arms, legs, or tails to enhance body capabilities (e.g., walk, run, or complete tasks faster)*
• *To control multiple avatars and overcome the limitations of a single physical body*
• *To experience new senses*
• *To act as a complementary body for people with reduced mobility (assist in tasks such as grocery shopping or cooking)*

*Expanded cognitive abilities*
- *To be able to interact with people in other languages*
- *To have new ways of learning or reasoning*

*Immortality*
- *To become immortal by transferring consciousness to the avatar*

**SM - 2**

Selection of representative quotes per thematic category from the question "If you could create the ideal robot or virtual avatar, which human capability would you like it to enhance?"

**Communication abilities**

- "Cross-cultural communication: The avatar could help people understand and navigate cultural differences, promoting empathy and respect in diverse communities. By enhancing human empathy, robots and virtual avatars could contribute to a more compassionate and understanding"

- "Language barrier to improve communication with different people with different languages"

- "To understand the emotions of a person that how he feels right now and the ways to improve his mental health"

- "If I could create the ideal robot or virtual avatar, I would like it to enhance human capability in the area of communication and understanding. This advanced avatar would have the ability to deeply understand human emotions, intentions, and context, enabling it to communicate effectively and empathetically with individuals from diverse backgrounds. It would be skilled in interpreting subtle cues, adapting its responses accordingly, and fostering meaningful connections with users. By enhancing this aspect of human interaction, the avatar would be able to provide personalized support, guidance, and companionship, ultimately making interactions more authentic, supportive, and enriching for everyone involved."

- "I would like the robot or virtual avatar to enhance communication skills, such as translating languages in real time, understanding emotions, and offering personalized advice or solutions. This would make it highly adaptable and useful in diverse environments."

- "Improving communication skills to facilitate clearer and more effective interactions with humans would be a valuable enhancement."

- "If I could create the ideal robot or virtual avatar, I would design it to enhance empathy and emotional intelligence in human interactions. This capability would allow it to: 1. Understand and Respond to Emotions: • Detect subtle emotional cues in a person's voice, facial expressions, or behavior. • Respond with tailored support, such as calming an anxious traveler at an airport or cheering up a frustrated shopper in a mall. 2. Bridge Communication Gaps: • Offer multilingual support to facilitate conversations between individuals who speak different languages. • Use gestures and expressions to communicate effectively in scenarios where words might not suffice. 3. Provide Personalized Assistance: • Recognize individual preferences and adapt its behavior accordingly (e.g., being more patient or humorous depending on the person's mood). • Remember past interactions to offer continuity and a sense of familiarity. 4. Foster Inclusivity: • Assist people with disabilities by…"

**Cognitive abilities**

- "If I were to invent a machine, I think it would be something that enhances human creativity and problem-solving abilities. Imagine a device that could instantly generate innovative ideas, artistic creations, or solutions to complex problems. It could work by analyzing vast amounts of data, combining different fields of knowledge, and presenting novel concepts or designs that humans might not have thought of on their own. This machine could be a tool for researchers, artists, engineers, and anyone else looking to push the boundaries of what's possible. It wouldn't replace human creativity but rather amplify it, acting as a catalyst for inspiration and breakthroughs. I imagine it as a collaborative partner, helping people explore new avenues of thought and discovery

- "If I could create the ideal robot or virtual avatar, I would want it to enhance human creativity. By amplifying creative thinking, this avatar could help people explore new ideas, solve problems in innovative ways, and unlock their full imaginative potential. For instance, in creative industries like design, writing, or art, the robot could collaborate with humans by suggesting novel concepts, offering alternative perspectives, or even helping to brainstorm solutions when facing creative blocks. It could assist in developing new stories, music, or artwork by analyzing patterns and offering unexpected combinations or prompts, encouraging individuals to think beyond conventional boundaries. In educational contexts, the avatar could foster a more creative approach to learning, encouraging students to explore subjects in unconventional ways, experiment with different solutions, and embrace trial and error as part of the learning process. By enhancing creativity, the avatar could help indiv…"

- "The ability I'd most like to enhance with a robot or virtual avatar is creativity. Creativity is a core part of what makes us human, allowing us to innovate, solve problems in new ways, and express ourselves artistically. However, it can be a difficult skill to develop and can sometimes be hindered by self-doubt or lack of inspiration."

- "Enhancing the ability to learn and adapt quickly to new situations would be beneficial for a robot or virtual avatar."

- "I would like the ideal robot or virtual avatar to enhance problem-solving skills, such as providing quick and effective solutions to complex issues, whether it's navigating an airport, resolving technical problems, or offering tailored advice in customer service"

- "The brain's capabilities (storing data), visuals, and recordings also Enhance decision-making by providing instant, data-driven insights and personalized recommendations."

- "If I could create the ideal robot or virtual avatar, I'd want it to enhance our ability to manage time more effectively. Imagine a robot that helps you organize your day, reminds you of important tasks, and even suggests the best times for breaks or activities based on your energy levels. That would be such a game-changer."

**Augmented body capabilities**

- "Super strength that can easily carry heavy objects."

- "The human capability I would like it to have is skin like structure but made from metal."

- "Agile speed, quick movement and reaction."

- "The robot who can handle emergency situations like saving life at fire scene"

- "The human capability that I like to enhance is mobility to access inaccessible areas such as a building on fire with full of harmful smoke."

- "The ability to never let the eyesight weaken and be able to multitask without breaking down"

- People who have disabilities robots can help - so eyesight or movement"

- "eyesight and mobility"

**Sensory capabilities**

- "Skin touch"

- "Smelling capabilities"

**Mental health**

- "I would like to make it sense how I will feel. If I am sad it makes me happy by jokes and all"

- "My mind and mental health"

**Work-related tasks**

- "Working"

- "Capacity to cook and do household chore"

**Super-hero inspired capabilities and altruistic behaviors**

- "This one I have will be more like BB8 from Star Wars which will understand my voice as authentication and do work only for me – there is no restriction to what it can learn, in one word "BB8 learn Karate", it learns in a moment with deep learning and all those stuff. Of course, I don't want my robot to be turned into an iRobot one (watch this Will Smith if you haven't) which gets consciousness and emotions."

- "If I could create the ideal robot or virtual avatar, i would like it to enhance the spider man characteristics"

- "A robot or virtual avatar that could help fight hunger and war."

**No wish to enhance capabilities through avatars**

- "None"